\documentclass[aps,a4paper,12pt,twoside]{revtex4}

\usepackage{epsfig}
\usepackage{epstopdf}
\usepackage{amsmath,amssymb,color,amsthm}
\usepackage{tikz}
\usepackage{color}
\usepackage{multirow}

\usepackage[english]{babel}

\parskip=\medskipamount

\newcommand{\fig}[1]{Fig.~\ref{#1}}

\newcommand{\be}{\begin{equation}}
\newcommand{\ee}{\end{equation}}

\newcommand{\barr}{\begin{array}}
\newcommand{\earr}{\end{array}}

\newcommand{\beqn}{\begin{eqnarray}}
\newcommand{\eeqn}{\end{eqnarray}}

\newcommand{\bw}{\begin{widetext}}
\newcommand{\ew}{\end{widetext}}

\newcommand\disp{\displaystyle}

\newcommand{\la}{\left<}
\newcommand{\ra}{\right>}

\def\runninghead#1#2{\pagestyle{myheadings}
\markboth{{\protect\it{\quad #1}}\hfill} {\hfill{\protect\it{#2\quad}}}}

\begin{document}

\runninghead{\sl  O.V. Valba, S.K. Nechaev, O. Vasieva}{On prediction of regulatory genes by
analysis of C.elegans functional networks}

\title{On prediction of regulatory genes by analysis of C.elegans functional networks}

\author{O.V. Valba$^{1,2}$, S.K. Nechaev$^{1,3}$, O. Vasieva$^{4}$}

\affiliation{$^{1}$LPTMS, Universit\'e Paris Sud, 91405 Orsay Cedex, France \\ $^{2}$Moscow
Institute of Physics and Technology, 141700, Dolgoprudny, Russia \\ $^{3}$P.N. Lebedev Physical
Institute of the Russian Academy of Sciences, 119991, Moscow, Russia \\ $^{4}$Institute of
Integrative Biology, University of Liverpool, Crown St., Liverpool, L69 7ZB, UK}

\begin{abstract}
Connectivity networks have recently become widely used in biology due to increasing amounts of
information on the physical and functional links between individual proteins. This connectivity
data provides valuable material for expanding our knowledge far beyond the experimentally validated
via mathematical analysis and theoretical predictions of new functional interactions. In this paper
we demonstrate an application of several algorithms developed for the ranking of potential
gene-expression regulators within the context of an associated network. We analyze how different
types of connectivity between genes and proteins affect the topology of the integral C.elegans
functional network and thereby validate algorithmic performance. We demonstrate the possible
definition of co-expression gene clusters within a network context from their specific motif
distribution signatures. We also show that the method based on the shortest path function (SPF)
applied to gene interactions sub-network of the co-expression gene cluster, efficiently predicts
novel regulatory transcription factors (TFs). Simultaneous application of other methods, including
only interactions with neighborhood genes, allows rapid ranking of potential regulators that could
be functionally linked with the group of co-expressed genes. Predicting functions of regulators for
a cluster of ribosomal/mRNA metabolic genes we highlight a role of mRNA translation and decay in a
longevity of organisms.
\end{abstract}

\maketitle

\section{Introduction}

Analysis of functional properties of biological systems requires the integration of both the
functions of their individual components and the properties of the components interactions.
Reconstruction of functional networks from the known pair-wise connectivity of biological molecules
offers systems level insights into complex biologic phenomena. Though widely used and taken for
granted the network-based approach requires very accurate choice of data to reflect a proper
balance between maximizing connectivity on one hand, and preserving the reliability of network
behavior on the other.

Direct genetic and regulatory interactions as well as indirect indicators of functional links
between proteins, such as co-expression, co-occurrence, gene-fusion, {\em etc}, are used to provide
a generic topology of a network. It is known that subnetworks based on genetic and physical
protein-protein interactions overlap only weakly \cite{landscape}, which perhaps indicates
different levels of functional organization within complex biological systems. However, it has been
also documented that indirect indications or proxies of functional relevance between proteins, such
as gene co-expression and genome co-localization are largely complementary and correlate well with
ontology-based protein groupings \cite{Mansour, Fagan}. It is not clear to what extent
connectivities of different types affect the general architecture of an integrated network
structure.

The gaps in experimentally-derived knowledge on regulatory and structural features of biological
systems can be filled to some extent by theoretical predictions. However, a greater understanding
of principles of connectivity between biological functions at different levels of cellular
organization is needed to avoid false extrapolations in proposed network structures or topologies.
There is thus an immediate need for a means of evaluating the generally used organism-specific
functional networks and understanding of their underlying properties. Here we show how the
statistical analysis of network connectivity can be used to predict new gene expression regulators.
We suggest that co-expression clusters can be easily identified as highly-connected islands within
the integrated network, and that these islands can be used to suggest new regulatory genes for
subsequent verification. For this, we propose a new application of modified statistical algorithm
\cite{Vohradsky}, based on so-called "shortest path function" (SPF) to rank the nodes that have
most effect on genes expression. This can be also applied to any explicitly defined group of genes.

We retrieved gene co-expression clusters from existing large-scale microarray data on heat shock
responses in C.elegans \cite{Gutteling} and projected them onto subnetworks composed from a
combination of different connectivity types (gene interactions, protein interactions, gene
co-expression) in WormNet database of pair-wise undirected functional links between
proteins\cite{WormNet}. Correlation between experimentally observed co-expression links and
connectivity of genes in Wormbase \cite{WormBase} is shown to be very strong especially for
ubiquitously expressed ribosomal, proteosomal and exosomal gene clusters. Via application of SPF
method we also predicted several transcription factors (TFs) as potential regulators of the
analyzed groups of co-expressed genes. By searching for the most connected regulatory nodes within
the neighborhood  of each cluster we also noticed a strong link between nonsense mediated decay
(NMD) as well as longevity-related genes and the highly connected clusters mentioned above. We
suggest that these functional connections may explain a dependence of adult life span on the
metabolism of polyamines. Interpretation of the organism-specific integral biologica networks and
prediction of protein complexes and genetic regulators from a network context may benefit greatly
from our study and the new algorithms.

\section{Methods}

\subsection{Data preparation}

Microarray data are adopted from \cite{Li} where two parental C.elegans strains N2 (Bristol) and
CB4856(Hawaii) and the recombinant inbred  strains were used to measure gene expression in a
control conditions and after a heat shock. N2 and CB4856, represent two genetic and ecological
extremes of C. elegans \cite{17,18, alda}. Their genetic distance amounts to about one polymorphism
per 873 base pairs \cite{19}. Both strains have contrasting behavioral phenotypes (solitary versus
gregarious) \cite{18} and differ strikingly in their response to a temperature change
\cite{Gutteling}. In \cite{Anderson} the strains were exposed to 16 $^0$C and 24 $^0$C,
temperatures that are known to strongly affect phenotypic characteristics such as body size,
lifespan, and reproduction \cite{Gutteling, Kammenga}.

Gene expression patterns were assessed by oligonucleotide microarray hybridization. All microarray
data have been deposited by the authors in NCBI's Gene Expression Omnibus (GEO, \cite{GEO}) and are
accessible through the GEO Series accession number listed under the Accession Numbers GSE5395.
Absolute expression values have been used for KMC clustering analysis by means of Mev4 application
\cite{Mev-4} with requested 50 clusters/Euclidian distance. The resultant clusters were used as
"co-expression" clusters in our analysis.

String software \cite{string,Szkl} has been used to reconstruct graphical networks from the sets of
C.elegans genes. WormBase database \cite{WormBase} has been used for ID retrieval and translation
of IDs into gene names, associated functional annotations and ontological categories. WormNet
\cite{WormNet} has been used a source of information on pair-wise interactions between genes. We
also used it to retrieve data on separated genetic interactions of C.elegans genes and
co-expression links in C.elegans.

\subsection{Methods of statistical analysis of network connectivity}

To investigate the connectivity properties of the gene clusters on WormNet and to establish
potential regulators for the gene clusters we use the following approach. Denote by $d_{i,M}$ the
shortest path along the network from a given vertex  (regulatory gene) $M$ to some other vertex
(cluster gene) $i$. Consider the shortest path function (SPF) determined as follows
\be
k^{SPF}_{M}=\frac{1}{N}\sum_{i=1}^{N}\frac{1}{d_{i,M}}
\label{eq:1}
\ee
where the summation is performed over all cluster genes ($N$ is the number of the genes in the
cluster, i.e. the cluster size). The connectivity of the gene cluster on the whole network and/or
on its subnetworks we describe by SPF defined for all cluster genes
\be
k^{SPF}_{cl}=\frac{2}{N(N-1)}\sum_{i,j=1}^{N}\frac{1}{d_{i,j}}
\label{eq:2}
\ee
here $N$ is the number of the genes in the cluster, $d_{i,j}$ is the shortest path between the
nodes $i$ and $j$. Thus defined, the SPF has a very transparent meaning, since it gives the
averaged reciprocal paths between pair of cluster genes. If $i$ and $j$ are not linked on the
network, the contribution to SPF from this pair $(i, j)$ equals to zero, while the maximal
contribution is reached for directly linked pairs. So, SPF can be used to characterize
quantitatively the connectivity of a gene cluster in a given network. For comparison we use also
another method for determination the cluster connectivity. The so-called "connectivity coefficient"
is defined as a ratio between the number of $inner$ links (which connect only cluster genes) to the
number of $all$ links which the cluster has in the network:
\be
f_{cl}=\frac{N_{in}}{N_{all}}
\label{eq:3}
\ee
We consider WormNet and its parts as unweighted undirected network. Note that SPF analysis can be
easily extended for weighted and directed networks as well.

Another approach to characterize a connectivity (and topology) of a subnetwork (cluster) is to
investigate its motifs distribution. A network is a set of vertices (or nodes) and connections
between them (links or edges) without self--connections and multiple edges. The network is random
if any link occurs independently on others with a certain probability. The network is directed if
any link either has an orientation $i \to j$, or is bidirectional $ i \leftrightarrow j$. Otherwise
the network is non-directed. The network is set by the adjacency matrix. The local topological
properties of networks, both directed and non-directed, for given number of vertices and vertex
degree distribution are characterized by the rates of connected subgraphs. Since the number of such
subgraphs grows combinatorially with their sizes, usually only small subgraphs are considered. In
particular, in the works \cite{Alon1, Alon2,Alon3} only subgraphs of size 3 and 4 have been
analyzed for directed and undirected networks.

The rates of subgraphs in a given network depend on the vertex degree distribution. This
complicates the comparison of networks of different sizes and of different degree distributions by
the rates of their subgraphs. In order to compensate these differences, the procedure of so-called
network randomization was proposed in works \cite{Alon2, Alon3}. In this procedure the network
experiences multiple permutations of links under the condition of conservation in each vertex the
number of incoming, outgoing and bidirectional links. Using this method an ensemble of randomized
versions of a given network is generated, and for every subgraph the statistical significance
\be
Z_{k}=\frac{N_{k}-\la N_{k}\ra}{\sigma_{k}}; \quad k=[1...m]
\label{eq:4}
\ee
is calculated, where $N_{k}$ is the amount of $k$-th subgraphs in the initial network;  and  $\la
N_{k}\ra$ and $\sigma_k$ correspondingly the mean and the standard deviation of subgraphs of given
type in the randomized networks; $m$ is the total number of considered subgraphs. Subgraphs with
the statistical significance essentially exceeding 1 are called motifs \cite{Alon1,Alon2,Alon3}.
The distribution of motifs in the network under consideration is characterized by the significance
profile which is a normalized vector $x=\left\{x_{1},....,x_{m} \right\}$ of statistical
significance of all subgraphs of given size. The components of the vector $x$ are
\be
x_{k}=\frac{Z_{k}}{\sqrt{\disp\sum_{k=1}^{m}Z_{k}^{2}}}; \quad k=[1...m]
\label{eq:5}
\ee

\subsection{Prediction of potential regulators for the gene clusters}

We use the SPF coefficients $k^{SPF}_{M}$, determined in \eqref{eq:1} for a search for potential
regulators of co-expressed gene clusters. Note, that potential regulators can belong to the cluster
or lie outside of it. For the search of the potential cluster regulators it is important that the
regulator is specific to it. It has many connections to the gene cluster but at the same time is
weakly connected to other genes in the network, which the cluster belongs to. In other words, the
potential regulator must have enough links to control the cluster not being herewith the hub of the
network. We introduce the value which takes into account the connectivity of the regulatory gene in
the whole network:
\be
K_{M}^{SPF}=\frac{\disp\sum_{i=1}^{N}\frac{1}{d_{i,M}}}{\disp\sum_{i=1}^{G}\frac{1}{d_{i,M}}}
\label{eq:6}
\ee
By definition, the $K_{M}^{SPF}$ is the ratio of the sum of the shortest paths from regulator to
the cluster genes, and the sum of the shortest path from regulator to all genes in the network. one
sees, that $K_{M}^{SPF}$ measures how specific is the regulator to the cluster. The function
\eqref{eq:1} can be considered as a first term in expansion of the function \eqref{eq:6}. We use
this value for the search of potential regulators to the gene clusters in the network and its
subparts. We compare this method with some other described below.

Namely, for any potential regulator we define the fraction of the cluster genes, which are
connected to it:
\be
f_{M}=\frac{N_{M}}{N}
\label{eq:7}
\ee
Analogously to \eqref{eq:6} we introduce the value, which takes into account the "interaction" of
the potential regulator with outer part of the cluster genes:
\be
F_{M}=\frac{N_{M}}{N_{M}(G)}
\label{eq:8}
\ee
Here, $N_{M}(G)$ is the number of all links, which the potential regulator $M$ has in the whole
network $G$.

\section{Results}

\subsection{Statistical properties of co-expression clusters}

\begin{figure}[ht]
\epsfig{file=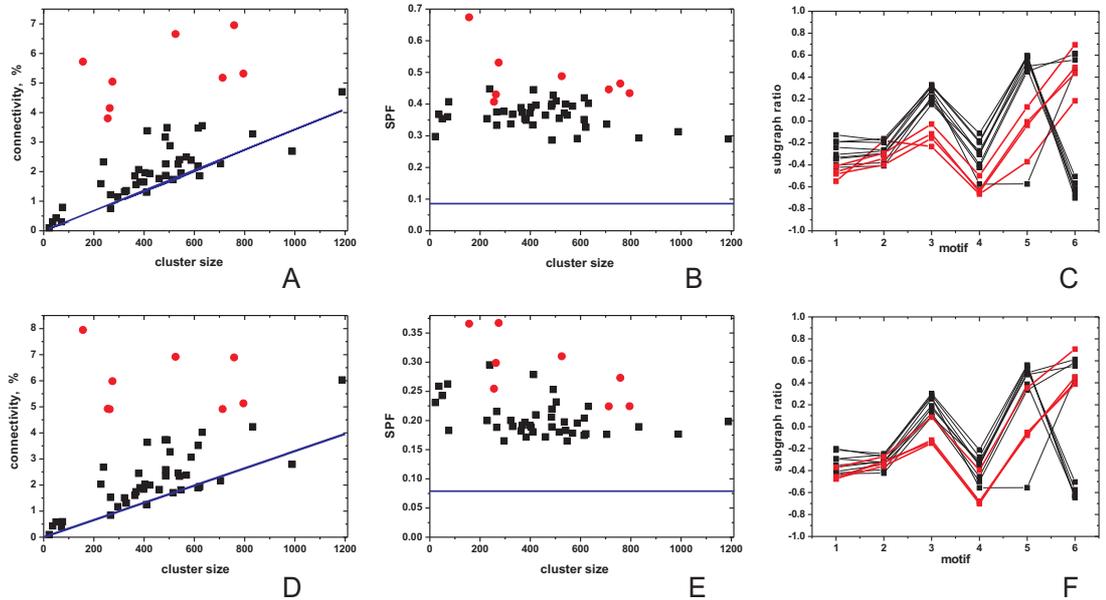,width=15cm} \caption{The connectivity coefficient (A and D), SPF (B and
E) and motif distribution (C and F) in the whole WormNet (up) and in its expression subpart
(below).}
\label{fig:2}
\end{figure}

Our first goal is the investigation of the properties of extracted gene clusters in the whole
network WormNet. The \fig{fig:2} presents the dependencies of the cluster's connectivity and SPF on
the cluster's size (the number of genes included in the cluster), as well as the motif's
distribution for some clusters. Connectivity coefficient is generally proportional to a cluster
size. In average, an expression cluster is composed by genes connected in WormNet stronger than any
group of randomly selected genes -- see \fig{fig:2}A. The SPF values for all gene clusters lie much
higher than the SPF coefficients for random clusters shown in the \fig{fig:2}B (here and below: a
random cluster is a cluster with randomly chosen nodes in a WormNet). The most connected clusters
are bounded mainly by co-expression links in the WormBase \fig{fig:2}D. We have selected the
clusters with very high connectivity coefficients. Such clusters are well-connected and are
depicted by red points in the graphs. These clusters also are characterized by different motif's
distribution with a prevail of linear 4-nodes chains \fig{fig:2}C. According to the motif's
distribution of gene clusters, we can divide them into two groups. The first group includes the
gene clusters, in which the fully connected subgraphs dominate and all well-connected clusters
belong to this group. The second group is formed by the clusters with a small number of fully
connected motifs. It is worth noting, that protein structure networks are also characterized by the
same motif distribution \cite{Alon2}.

The number of co-expression and physical connections strongly prevail over genetic interactions in
the WormBase, which emphasizes the necessity in additional experimental analysis, or in-situ
prediction of potential regulatory modules for C.elegans. Our analysis shows that the motif's
profiles for expression sub-network of gene clusters are very close to the profiles in the whole
WormNet \fig{fig:2}C,F. It means that in our clusters the co-expression links prevail over other
links. This observation provides a proof that genes of well-connected clusters are mainly connected
to each other by co-expression. Also, our analysis demonstrates that expression links are
responsible for the existence of the cycles in well-connected clusters \fig{fig:2}F.

\subsection{Prediction of expression cluster regulators}

We have used the most connected cluster 1 to find an optimal algorithm to predict potential
regulators of mRNA pool for the groups of co-expressed genes. Two different methods have been
tested.

The first method is based on ranking the network nodes according to a fraction of cluster genes
(FCG) they are directly connected to. This method requires dense connectivity matrix to be
efficient. The predicted functions are expected to have a strong and specific involvement in
regulation of a particular co-expressed gene cluster. However, the ranked gene list needs filtering
to eliminate a malfunctioning.

The second method is based on ranking nodes according to the shortest average distance to all
cluster genes (SPF method). As in the first method, the obtained gene lists needs filtering. From
the \fig{fig:2} one can see that the clusters have different topologies in different connectivity
subnetworks, with the genetic interactions subnetwork being the least connected. The power of the
SPF method is that it compensates the absence of knowledge about regulation of many genes and
quantitatively validates the probability of co-regulation of a group of genes by different nodes.
SPF method does not require a matrix to be dense and can be applied to a subnetwork of genetic
interactions producing the list of regulators without filtering. FCG and SPF methods are applied to
a total network and to the co-expression connectivity subnetwork with similar outcomes. The top of
the predicted regulators is presented in Table ~\ref{FCG}.

\begin{table}[ht]
\begin{tabular}{|l|l|l|} \hline
\textbf{Seq. IDs} & \textbf{Gene} &\textbf{ Function}\\ \hline
F57B9.6 & inf-1 & Transl.initiation/ RNA transport  \\
\hline
T05G5.10 & iff-1 & Transl.initiation/ NMD \\
\hline
Y71G12B.8 & Y71G12B.8   &  RNA helicase/ RNA transport \\
\hline
T10C6.14, T10C6.12, T10C6.11, F45F2.3, &  \multirow{10}{*}{38 His genes} &  \multirow{10}{*}{Histones}\\
 F45F2.4, F45F2.12, ZK131.4, ZK131.6, & & \\
 ZK131.8, ZK131.10, K06C4.10, K06C4.11,& &  \\
 K06C4.4, K06C4.3, K06C4.12, ZK131.1, & & \\
 K06C4.2, F35H10.1, F17E9.12, F17E9.13, & & \\
  C50F4.7, K03A1.6, C50F4.5, F08G2.2,& & \\
  B0035.9, B0035.7, F07B7.9, F07B7.10, & & \\
  F07B7.4, F07B7.3, F07B7.11, F54E12.3, & & \\
  F54E12.5, F55G1.11, F55G1.10, F22B3.1, & & \\
 H02I12.7, T23D8.5, T23D8.6 & &     \\ \hline
C41D11.2     &eif-3.H & Transl.initiation\\ \hline
F32E10.1 &  nol-10 &    nucleolar protein, polyglut. binding\\ \hline
F54H12.6     &eef-1B.1 &    Elongation factor\\ \hline
C01F6.5 &    aly-1   &RNA export\\ \hline
M163.3   &his-24     &Histones\\ \hline
B0564.1 &   tin-9.2  &decay/ NMD\\ \hline
Y18D10A.17 &    car-1 & decay/decapping\\ \hline
F56D12.5     &vig-1  &RISC component/miRNA binding\\ \hline
F26D10.3 &  hsp-1 & splicing\\ \hline
R04A9.4  &ife-2  &Transl.initiation\\ \hline
\end{tabular}
\caption{Top of the predictable regulators for test cluster N1 by FCG method}
\label{FCG}
\end{table}

Several predicted regulators are well integrated both in the top cluster networks and in the
network reconstructed from the genes with a longevity phenotype retrieved from WormBase. Among the
most promising predicted regulators that could connect the top clusters with a longevity regulation
are: daf-2, iff-1, cgh-1, tin9.2, car-1. They all are related to mRNA processing/translation/decay
and are in a cross-talking relationship (Table \ref{FCG}). According to their position in the
network, they may play a role of linkers between top connected co-expression clusters related to
ribosomal biogenesis, proteosome and central metabolic functions (Fig 1). The SPF method has also
been applied to the gene regulatory network. The results of this analysis (the top ranked
regulatory nodes) are summarized in the Table \ref{SPF}. Note that the SPF method allows us to
predict some regulators which we could not detect by the FCG method, and the power of this method
had been enforced by its application to a gene regulatory connectivity subnetwork, as it is clearly
demonstrated in Table \ref{SPF}. We could identify a number of TFs that may be considered for a
role of transcription regulators of genes in the top co-expression cluster 1, such as: F30F8.8,
transcription initiation factor TFIID subunit 5; R74.3, xbp-1, heat-shock transcription factor;
F02E9.4, sin-3-histone deacetylase subunit; 33A8.1, pre-mRNA-splicing factor CWC22. This method
also greatly increased the ranking position of daf-2 and genes upstream daf-2 (C25A1.10) or being
directly affected by daf-2 mutation (C05C8.3) an immediate potential connection to a group of genes
with longevity phenotype that have a strong overlap with our cluster 1.

\begin{table}[ht]
\begin{tabular}{|l|l|l|} \hline
\textbf{Seq. IDs} & \textbf{Gene} &\textbf{ Function}\\ \hline
Y55D5A.5,B0334.8,Y116F11B.1 & daf-2, age1, daf-28 & insulin/aging \\ \hline
F35H8.5 & exc-7 & mRNA processing \\ \hline
W10D5.1 & mef-2 & TF\\ \hline
C17D12.2    & unc-75    & Splicing\\ \hline
C47G2.2 & unc-130   & TF\\ \hline
F30F8.8 & TFIID & Transl.initiation\\ \hline
R74.3   & xbp-1     & TF, histone modulation\\ \hline
F33A8.1 & cwc22 & Splicing\\ \hline
C41C4.4 & xre-1 &  (RNA processing) decay/ processing\\ \hline
C37H5.8 & hsp-6 & Decay\\ \hline
C26D10.2    & hel-1 (helicase)  & DNA helicase\\ \hline
C07H6.5 &cgh-1 (decapping) &    decay/ decapping\\ \hline
F02E9.4 & sin-3 (HDAC)&     histone modulation\\ \hline
M163.3  & his-1 & Histone\\ \hline
C25A1.10    & dao-5 & rRNA transcription/aging\\ \hline
ZC247.3 & lin-11    & TF\\ \hline
R107.8 &    lin-12  & TF\\ \hline
C05D9.5 & ife-4 & Transl.initiation\\ \hline
R11E3.6 & eor-1 & TF\\ \hline
F43G9.11    & ces-1 & TF\\ \hline
ZK909.4 & ces-2 & TF\\ \hline
\end{tabular}
\caption{ Top of the predictable regulators for test cluster 1 by SPF method on genetic subnetwork}
\label{SPF}
\end{table}

Here we do not consider exclusion of the hub-regulators by the procedures \eqref{eq:6} for SPF
method and \eqref{eq:8} for FCG method. Our statistical analysis demonstrates that the most
connected components of WormNet are included in the test cluster $No$1, so the methods \eqref{eq:6}
and \eqref{eq:8} give about the same list of potential regulators as FCG and SPF method
respectively; the ranging order is slightly changed, for example, in a green frame in the figure we
show gene (exc-7) that has been excluded in our SPF-hub-exclusion algorithm because its
connectivity exceeds a fixed threshold.

\begin{figure}[ht]
\epsfig{file=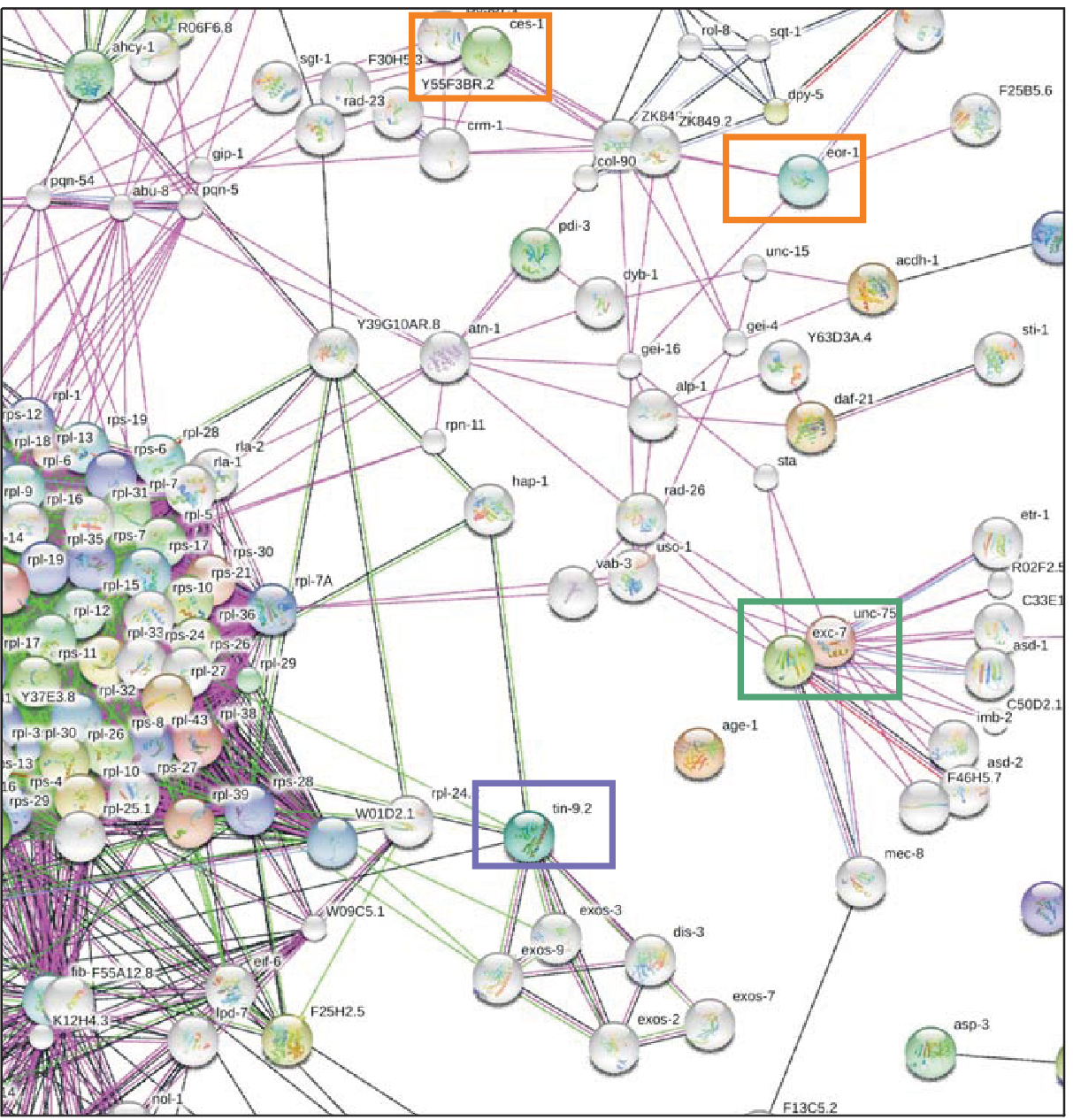,width=13cm}
\caption{Connectivity between the predicted regulators and Cluster 1 in STRING Network browser.
Evidence view for high confidence (0.700) connections. Pink connectors - experimentally derived
interactions, black-co-expression, and green-co-localization in genomes, blue-co-occurrences in
genomes. Colored circles-input genes, white circles-the most associated additional nodes (set
number of 200) automatically added by a STRING software on a request to increase a connectivity
between uploaded functions. Predicted potential regulators are shown in frames: orange-SPF method,
purple-FCG method, green node excluded in hub-exclusion SPF method.}
\label{net}
\end{figure}

The \fig{net} illustrates a typical position of the predicted potential regulators for the Cluster
1. Nodes predicted by FCG method (purple frame are proximal to the cluster or even inside the
cluster. The nodes predicted by SPF method may be significantly distant from the many nodes in the
cluster (ces-1, eor-1, orange frames on \fig{net}). Though the connections between the
SPF-predicted node and the cluster may include several intermediate steps, the majority of these
steps do contain the nodes that can translate signals at the level of mRNA pool regulation,
potentially representing complexes of proteins with a joint regulatory performance. As one sees,
the type of connectors utilized by different algorithms differs essentially: experimental,
regulatory connections are fundamental for SPF and more dense, co-expression ones, for FCG.

\section{Discussion}

Characterizing the degree of connectivity for a given gene to a specific set of genes in the
network as a normalized sum of inverse distances in a network, is very natural and brings us back
to the applications of "harmonic means" for graph analysis . The main conjecture behind the
application of harmonic mean to networks is as follows; if a vertex has multiple links to other
vertices, the information is sent "in parallel", i.e. concurrently along the network. Thus, one can
define the "efficiency", $e_{i,j}$ in communication between vertices $i$ and $j$ as the inverse of
the shortest distance, i.e. $e_{i,j}=1/d_{i,j}$, see \cite{Latora}. The average efficiency is
straightforwardly related to the definition of SPF. It should be noted that the interpretation of
the SPF function as the efficiency of communication could be very useful in further dynamic
analysis of the networks. Actually, let $v$ be the velocity, with which the information travels
along the network, then the amount of information sent from the node $i$ to the node $j$ per unit
time is just $v/d_{i,j}$ . The performance, $P$,  is the total amount of information propagating
over the network per unit time \cite{Latora}. In our forthcoming works we plan to analyze the
clusters taking into account their limited speed of information propagation. The concept of
performance seems very appropriate for that.

Well-connected clusters of co-expressed genes described in this paper largely represent protein
functional complexes, and they can be distinguished by the presence of a specific well-connected-6
link (unoriented) motif. This highly-connected motif can be used for detection of protein
functional complexes (islands) in integral networks. These islands, in turn, serve in prediction of
new regulatory nodes. In this study we used gene clusters derived from gene absolute expression
values data that probably increase detection of true protein complexes expressed from indeed highly
co-regulated genes \cite{GAV}.

Among the most interconnected clusters are the ones for ribosomal proteins and the regulation of
translation, proteosome, respiratory complex 1 and several central metabolic functions. Using the
most interconnected cluster 1 we tried to detect potential regulators from the associated network
context. Due to a non-directional nature of the edges in WormBase and an absence of directions in
the co-expression network we were unable to distinguish between a cause, consequence or undirected
physical interaction in a connected pair of proteins/functions and may only suggest the presence of
the functional linkage between the expressed genes and the regulators. However, additional data
from literature mining will likely help to vectorize the predicted interactions.

The cluster, which we used for the method validation, contains a large number of genes involved in the
translational machinery as well as several genes with a central metabolic role, among which we found
a large number of genes associated with a longevity phenotype in WarmBase database (Table
\ref{FCG}). The protein translation processes indeed have been recently considered for a central role
in the regulation of aging processes \cite{Syntich,Hamilton} and we assumed that the regulators
predicted in this study may also be linked to the processes underlying aging and the control of longevity.

The role of regulation of translational machinery by the insulin pathway in aging has been widely
discussed in literature \cite{Dillin, Kenyon}, however, the regulatory modules affecting the
expression of the related genes downstream of daf-2 have not been clearly defined. The candidates
suggested by the FCG algorithm, iff-1 and bir-2 were shown to depend on daf-16-insulin response
\cite{Murphy} and iff-1 has also been detected in gene expression screen for the longevity
phenotype in C/elegans \cite{Hamilton}. iff-1 is a eIF-5A homolog \cite{Hanazawa}, and eIF-5A that
links processes of mRNA translation to the nonsense-mediated mRNA decay (NMD) \cite{Shradera}.
Activation of eIF-5A requires posttranslational modification of one of the protein's lysines into
hypusine, and the enzyme that catalyses the first step of this modification, deoxyhypusine
synthase, performs the NAD-dependent oxidative cleavage of spermidine \cite{Myung}. Spermidine is
known to be involved in life span regulation and reproduction in a range of different organisms
\cite{Matt,Ravindar,Eisenberg}, though its mechanism of action is not clear. We suggest that its
stimulatory role in NMD via regulation of eIF-5A may be of importance in regulating translation and
as a consequence the life span of an organism. Interestingly, the ribosome maturation as well as
the mRNA binding SBDS protein \cite{Menne,Vasieva} that is linked in a network to iff-1 and tin9.2,
are both required for the longevity phenotype of daf-2 \cite{Samuel}.

Our analysis points to a potential role of mRNA decay processes downstream of the insulin-dependent
pathway in regulating translation and longevity. We suggest that elements of translational
machinery, that are regulated via insulin/caloric restriction, may be indeed non-responsive to
temperature changes as we see on the example of the Cluster 1 genes. Such persistence of expression
may require specific mechanisms of adjustment to altered kinetics of biochemical reactions and may
indeed involve a regulated mRNA decay process.  Homeostasis of pathways regulated by nutrients
supply regardless the temperature may lead to a very species-specific dynamic of cell survival and
growth and an organism`s life span adapted to the specific ecological dynamics of nutrients flow.
The transcription factors predicted in our study by SPF method may also occur to be involved in
aging and regulation of longevity . The genes cgh-1 \cite{Navarro}, dao-5 \cite{Simmer}, hel-1
\cite{Halas} were already linked to aging processes downstream daf-2, daf-16, and in case of dao-5
-- to a daf-16 independent pathway associated with determination of the adult life span GO-term in
WormNet database. Analysis of genotype-phenotype relationships \cite{Lehner} when more data for the
listed genes are available would allow deeper understanding of the direction of the defined links
and more narrow prediction of their function.

This work provides some new insights to the structure of biological functional networks and
highlights the aspects that need to be considered in prediction of regulatory nodes, protein
complexes and regulatory modules from a multilevel network context. We hope that it could be useful
in application to analysis of other organism`s networks and for improvement of analytic methods and
software in the relevant fields.

\begin{acknowledgments}
The authors are very grateful to Andrew Cossins for useful comments and to  L. Basten Snoek and Jan
Kammenga for providing us with the experimental data. The work is supported by ERASYSBIO Grant and
is a part of the GRAPPLE project.
\end{acknowledgments}

\end{document}